\documentclass[letterpaper]{article}
\usepackage{aaai2027}
\usepackage[hyphens]{url}
\usepackage{graphicx}
\usepackage{natbib}
\usepackage{caption}
\usepackage{booktabs}
\nocopyright

\title{OdinEval: A Reproducible Benchmark for LLM-Based Program Repair in the Odin Programming Language}
\author{
Bang Xie\textsuperscript{\rm 1}, \quad Hao Liu\textsuperscript{\rm 1}, \quad
Zhiyuan Peng\textsuperscript{\rm 1}, \quad Xin Yin\textsuperscript{\rm 2},\\
Senjian Zhang\textsuperscript{\rm 1}, \quad Yuan Luo\textsuperscript{\rm 1}, \quad
Chenhao Ying\textsuperscript{\rm 1}\corresponding, \quad Haiming Jin\textsuperscript{\rm 1},\\
Wei Chen\textsuperscript{\rm 1}, \quad Shaocong Long\textsuperscript{\rm 1}, \quad
Zhenyu Shi\textsuperscript{\rm 1}
}
\affiliations{
\textsuperscript{\rm 1}Shanghai Jiao Tong University, Shanghai, China\\
\textsuperscript{\rm 2}Zhejiang University, Hangzhou, China\\
yingchenhao@sjtu.edu.cn
}

\begin{document}
\maketitle

\begin{abstract}
Repository-level repair benchmarks still center on a few mainstream languages, leaving systems languages such as Odin largely untested. We present OdinEval, a reproducible benchmark built from documented defects in public Odin repositories. Each instance binds an issue to base and fix commits, a gold patch, an issue-specific regression test, a historical toolchain, and execution records. Admission requires the test to fail on the base revision and pass after the gold fix. When no usable developer test exists, a black-box test is reviewed independently by three instances of the same model, executed in both historical states, and revised from recorded feedback under a versioned Test Writing Skill. We evaluate six language models on 168 filtered instances under one shared protocol. Kimi-K3 records the highest Resolved score at 66.7\%, while Qwen3.8-Max has the highest Repro score at 96.4\%. The release includes frozen data, source archives, containers, validators, model patches, and audit manifests.
\end{abstract}

\section{Introduction}

Repository-level program repair turns an issue report and a historical codebase into a patch that builds and restores the reported behavior. Although LLMs are widely studied in software engineering \citep{zhang2024survey}, executable repair benchmarks mainly cover Java and Python, including Defects4J, BugsInPy, and SWE-bench \citep{just2014defects4j,widyasari2020bugsinpy,jimenez2024swebench}. Multilingual and ecosystem-specific benchmarks broaden this coverage \citep{zan2025multiswebench,peng2025soleval,wang2026arkrepobench,xie2026arkeval}, but none tests repair in Odin. It is therefore unclear whether current methods transfer to Odin's toolchains, project layouts, and test conventions.

Odin repositories provide an underrepresented systems-language repair setting. A benchmark for this setting needs more than a commit diff: it must identify the defective revision, reconstruct its toolchain, attach a behavioral test, and show that the test distinguishes the buggy and fixed states. Without this evidence, a patch may appear successful because the test did not run, the environment changed, or the assertion copied an implementation detail.

We introduce \textbf{OdinEval}, a benchmark and protocol for issue-linked Odin repairs. Its construction adapts ArkEval's auditable workflow \citep{xie2026arkeval} to Odin-specific mining, toolchains, and test execution. Each repair instance is a versioned evidence bundle. We prefer developer regression tests; otherwise, an iterative workflow generates a black-box test, collects three isolated same-model reviews, runs the test in both historical states, and records each revision of the test and its governing Skill.

This paper makes the following contributions:
\begin{itemize}
    \item \textbf{An executable Odin repair benchmark.} Each accepted instance will connect an issue, base and fix commits, gold patch, issue-specific test, historical toolchain descriptor, and immutable execution evidence.
    \item \textbf{A test-oracle construction protocol.} The protocol admits an instance only when the added test fails on the base revision and passes after the gold fix under the same declared environment; it separates runtime evidence from infrastructure failures.
    \item \textbf{A controlled LLM repair study.} We compare six models on the same 168 filtered instances while holding the repair prompt, source scope, validator, and reporting taxonomy fixed.
    \item \textbf{Reproducible research artifacts.} OdinEval will release frozen data records, source archives, containerized environments, validators, model patches, and audit manifests.
\end{itemize}

The current model comparison covers the 168 instances that passed benchmark construction and filtering.

\section{Background and Scope}

\subsection{Repository-Level Repair Benchmarks}

Defects4J \citep{just2014defects4j}, BugsInPy \citep{widyasari2020bugsinpy}, and SWE-bench \citep{jimenez2024swebench} connect defects to executable repositories and tests. Textual similarity alone misses patch application, buildability, test discovery, and behavioral outcomes. Repository-level code benchmarks likewise show that project definitions, APIs, and dependencies are part of the task \citep{yu2024codereval,li2024deveval}.

Existing collections expose different kinds of evidence. Some package curated defects with test suites; others bind natural-language issues to repository snapshots. Neither guarantees that a new issue-specific test separates the buggy revision from the fix, especially when projects rely on old compilers, local scripts, or unusual test entry points. OdinEval therefore releases the executable evidence with each instance.

\begin{figure*}[t]
\centering
\includegraphics[width=\textwidth,trim=0 1.15in 0 0,clip]{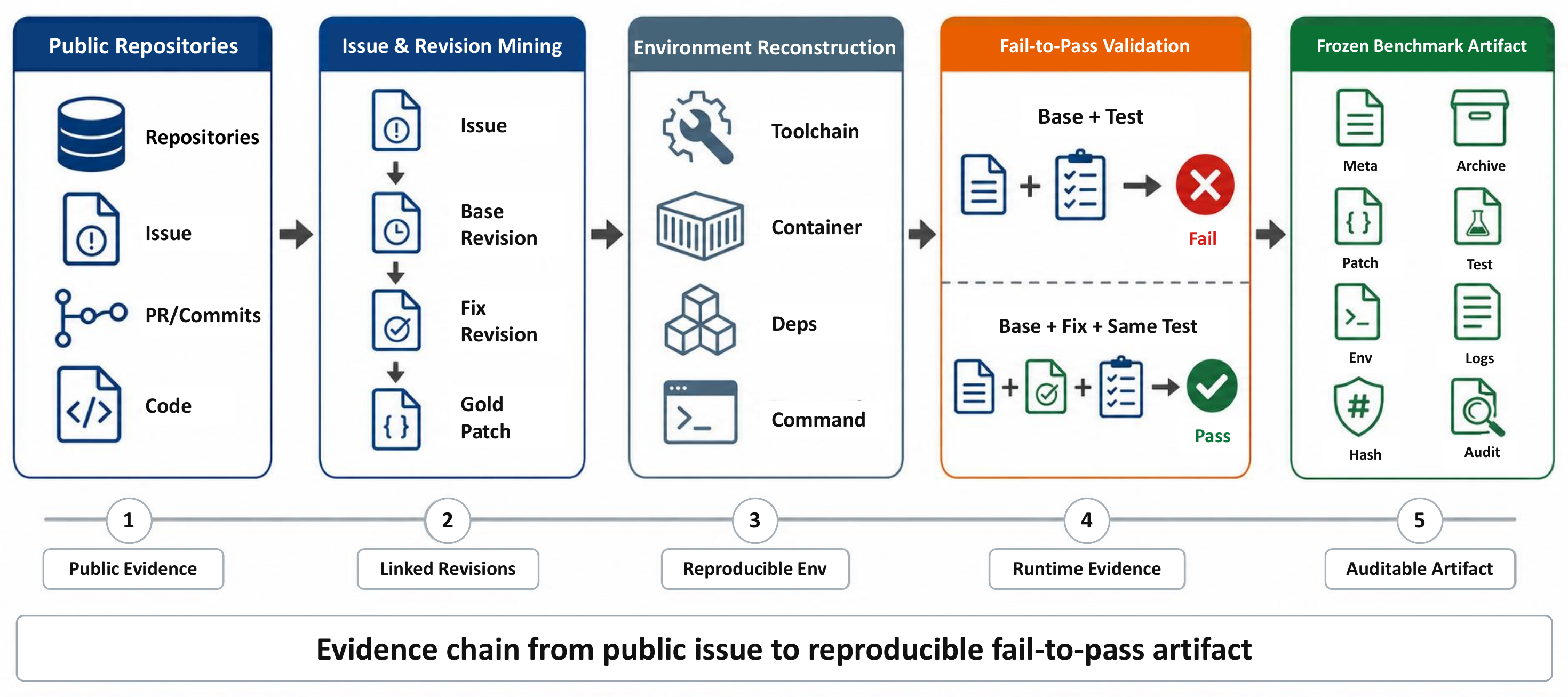}
\caption{OdinEval benchmark construction.}
\label{fig:benchmark-construction}
\end{figure*}

Figure~\ref{fig:benchmark-construction} summarizes the construction path: each stage adds evidence needed to make the final benchmark artifact independently auditable.

\subsection{Odin as an Evaluation Setting}

Odin is a systems programming language with a data-oriented design and an explicit toolchain \citep{odinlang}. Its public projects are a useful transfer setting because a repair system must operate over repository conventions that need not resemble the dominant Python- or Java-centric benchmark ecosystems. In particular, repository-level evaluation must preserve how a historical checkout obtains the compiler, resolves project-local dependencies, invokes tests, and distinguishes a test failure from an environment failure. A patch that is syntactically plausible but cannot be evaluated under the relevant compiler and command contract is not a completed repair.

Odin has no universal build layout or test framework. Each instance instead declares the smallest executable contract needed to reproduce its behavior. This supports heterogeneous repositories without hiding project-specific assumptions in a benchmark-wide script.

\subsection{Scope}

OdinEval is an \emph{issue-resolution} benchmark. It targets localized, behaviorally identifiable changes in public repositories; it does not claim to represent all Odin development work, large redesigns, or performance-only changes. The benchmark may include bug fixes and narrowly scoped corrections that are documented through an issue or equivalent project record. Each record must have a traceable base revision, a fix revision, and a reproducible execution path. The evaluation does not expose the gold patch, the fixed revision, or the gold test source to the repair model.

The unit of analysis is an accepted instance, not an issue label or a commit alone. An issue may map to multiple commits, and one commit may resolve multiple documented symptoms. OdinEval records these relations explicitly and includes a candidate only when curators can state one repair target and one validator contract. The benchmark excludes large-scale migrations, formatting-only changes, version bumps, and changes whose expected behavior cannot be observed under a reproducible command. These exclusions improve interpretability but also limit the scope of the conclusions; the final dataset report will disclose all exclusion categories and their counts.

\section{Benchmark Construction}

\subsection{Instance Contract and Admission Rule}

An OdinEval instance is a tuple
\[
I = (q, r_b, r_f, p_g, t, e, m),
\]
where $q$ is the documented issue, $r_b$ and $r_f$ are the base and fix commits, $p_g$ is the gold patch, $t$ is an issue-specific regression test, $e$ is a historical toolchain and execution specification, and $m$ is an audit manifest. The manifest records commands, container digest, return codes, test output, reviewer decisions, and checksums for released files.

An instance is admitted only when the same validator establishes:
\[
\mathrm{Run}(r_b + t, e) = \mathrm{FAIL}
\quad\mbox{and}\quad
\mathrm{Run}(r_b + p_g + t, e) = \mathrm{PASS}.
\]
The base failure must be attributable to the documented behavior, rather than to patch application, dependency resolution, compilation, test discovery, timeout configuration, or an unrelated failing test. Those conditions are preserved as diagnostic records but never counted as fail-to-pass evidence.

Passing a test suite does not by itself establish semantic correctness of a patch \citep{qi2015plausibility}, and test-oracle adequacy remains a separate challenge \citep{barr2015oracle}. OdinEval therefore retains explicit two-state evidence and oracle provenance rather than treating a passing outcome as a correctness proof.

\subsection{Repository Mining and Curation}

We will mine public Odin repositories with issue or pull-request histories. Repository inclusion requires public accessibility, substantive Odin implementation, revision pairs that can be retrieved, and an executable or reconstructable historical toolchain. The collection process has two deliberately separate stages: automated profiling produces a complete candidate record, and semantic curation decides whether that record represents an admissible issue-resolution task. The final paper will report the collection cutoff date, source repositories, candidate attrition, and the exact curation rubric rather than infer these quantities from the released benchmark alone.

\subsubsection{Candidate Profiling}

For every candidate, a profiler records the repository URL, issue and pull-request identifiers, available text, base and fix revisions, commit-parent relation, changed paths, added and removed lines, and any test or build commands found in repository metadata. The profiler also records failed retrievals rather than omitting them silently. This produces a candidate manifest before human judgment and makes it possible to distinguish selection decisions from later environment or test failures.

Patch size and the ratio of Odin source files are descriptive signals, not stand-alone admission rules. They help curators prioritize localized changes, but a small patch can still be a refactoring and a larger patch can still expose a single reproducible behavior. The final release will publish the exact counting definitions, including whether generated files, vendored files, and test-only paths contribute to each statistic.

\subsubsection{Semantic Curation}

Curators apply a written rubric to the profiled records. A candidate must have (1) a documented symptom or target behavior, (2) a traceable relation between that behavior and the historical change, (3) an executable base checkout or a credible reconstruction path, and (4) a feasible issue-specific oracle. They reject records whose semantic target is ambiguous, whose change is primarily a migration or broad redesign, or whose required dependencies cannot legally and reproducibly be obtained.

The curation record contains the decision, rationale, curator identifier or anonymous role, and any disagreement. If multiple curators are used, the final paper will report the independent-decision protocol, agreement statistic, reconciliation procedure, and final admission rule. No aggregate agreement value will be reported until it is computed from the frozen curation records.

\subsection{Historical Environment Reconstruction}

The evaluation environment is part of the benchmark definition. For each candidate that survives curation, OdinEval resolves the compiler release, operating-system base, dependency acquisition method, environment variables, working directory, test entry, filter, timeout, and expected exit code. These fields are materialized in a container recipe and a machine-readable command specification. The validator checks the revision before and after adding the gold patch using the same image digest and command sequence.

Environment reconstruction proceeds conservatively. A modern compiler is not silently substituted for a historical one, and a repository-wide command is not replaced with a narrower command merely to obtain a pass. When an exact historical toolchain cannot be reconstructed, the instance remains outside the accepted corpus. The release will retain the failed reconstruction manifest so that the selection boundary is inspectable.

\subsection{Regression-Test Construction}

When a developer test directly exercises the reported behavior, OdinEval adapts and validates that test against the two historical states. Automated test generation has long pursued executable tests through feedback-directed random testing and search-based generation \citep{pacheco2007randoop,fraser2011evosuite}; recent work also uses pretrained language models to augment this process \citep{lemieux2023codamosa}. When no usable developer test exists, the benchmark uses a versioned Test Writing Skill. The writer receives the issue, base repository, permitted test locations, and permitted execution commands. It must formulate an observable black-box oracle and must not assert source strings, private implementation details, exact imports, or the structure of the gold patch.

\subsubsection{Developer Tests and Test Adaptation}

Developer tests are preferred because they are closest to the repository's original maintenance intent. However, a test added in the fixed revision may depend on surrounding changes and therefore may not execute when transplanted to the base revision. OdinEval records whether a test is developer-provided, adapted from a developer test, or generated. An adapted test is accepted only after the same two-state validation used for generated tests. The benchmark never treats the mere existence of a test file in the fix revision as evidence of a valid oracle.

\subsubsection{Skill-Guided Black-Box Test Generation}

The Test Writing Skill defines the writer's allowed context, output format, and behavioral constraints. Before editing, the writer must state the issue trigger, observable behavior, and test command. White-box checks over source text, AST shape, visibility, exact identifiers, or signals from the official patch are forbidden. Each Skill version is immutable, and its change log links recurring rejections to the rules added in response.

Three isolated instances of the same reviewer model inspect each candidate test independently. Their rubric asks whether the test triggers the issue, observes the intended behavior, explains the base failure and fixed pass, and avoids implementation leakage. Unanimous approval is required before execution. Isolated same-model review is a consistency gate, not evidence of inter-model consensus; the artifact records the model identifier, prompt revision, and all reviewer rationales.

\subsubsection{Two-State Validation and Rewrite Loop}

An approved candidate then undergoes two-state runtime validation under the same declared environment. The validator executes the test first on the base state and then on the base state plus gold repair. It records command lines, output, exit status, elapsed time, and checksums of the checked-out source and applied test. Rejected tests are rewritten from structured feedback and re-enter all review and execution gates. A rejection caused by test discovery, build configuration, or toolchain setup is labeled separately from a behavioral mismatch.

The protocol does not claim that test generation or execution alone yields a complete oracle. It records the intended behavioral claim and two-state evidence because oracle adequacy remains a distinct testing problem \citep{barr2015oracle}.

The workflow may include a final human semantic audit. If this gate is used in the final corpus, the paper will state the auditors' role, whether they are blind to gold-patch details, the agreement rule, and the stored record. The human audit will not be described as completed until these fields and its evidence are available.

\begin{figure*}[t]
\centering
\includegraphics[width=\textwidth,trim=0 0.85in 0 0,clip]{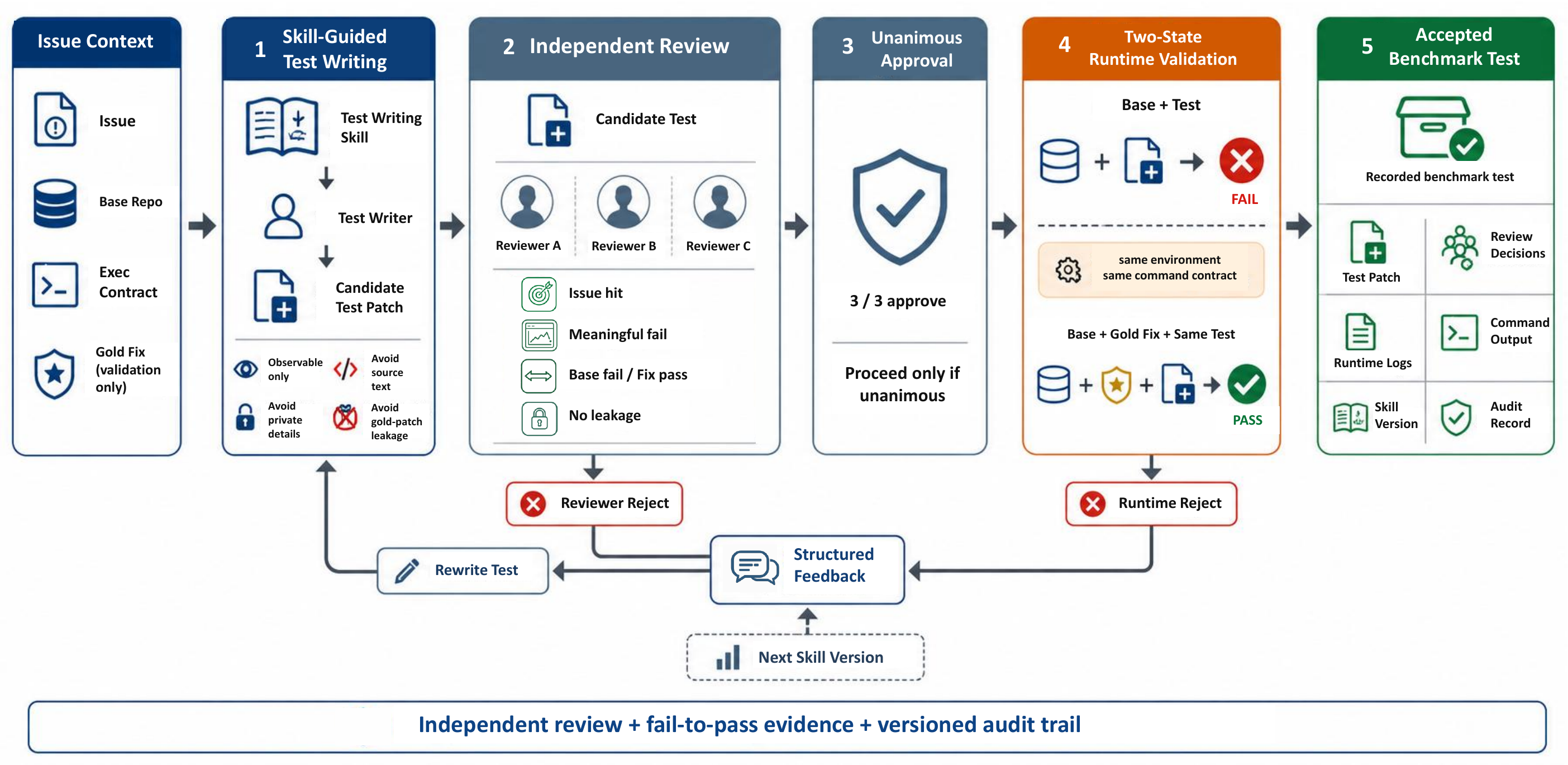}
\caption{Generated-test workflow.}
\label{fig:test-construction-workflow}
\end{figure*}

Figure~\ref{fig:test-construction-workflow} depicts this gated construction loop, including the retained feedback that links every rejection to a rewritten candidate test or a subsequent Skill version.

\subsection{Dataset Release}

The release will freeze both the dataset and the means to audit it. For each instance, OdinEval will provide a metadata record, source archive or retrieval recipe, containerized environment, validator, golden patch, test patch, and execution logs. Release checksums and a dataset-level manifest will bind the published results to a specific artifact version. The filtered corpus contains 168 instances.

The release checker verifies that every instance has all required files, that manifest hashes match their contents, that the two-state logs identify the same command contract, and that no generated-test record omits its review history. Dataset statistics, paper tables, and analysis scripts must read only this frozen manifest. The rule keeps an evolving working directory separate from the benchmark version used for reported results.

\section{Repair Evaluation Protocol}

\subsection{Task Formulation}

Given a base repository $r_b$, issue description $q$, and a localized source scope $L$, a model produces one unified patch $p$. The evaluator applies $p$ to $r_b$ and then runs the instance validator. A strict successful repair requires patch applicability, successful build or test preparation when applicable, passage of the issue-specific regression test, and passage of the declared non-target regression checks. We report every gate separately because a buildable patch is not necessarily a behavioral repair.

The repair model sees only information available at the base revision: the issue record, files in $L$, repository metadata permitted by the protocol, and the shared patch instruction. It does not see the fixed revision, the gold diff, the gold test patch, base-fail/fixed-pass logs, or reviewer evidence. The evaluator may use these withheld artifacts only after patch generation to score the attempt. This separation makes it possible to attribute a success to the model's generated patch rather than to leakage from the benchmark construction process.

Every generated patch is required to be a unified diff that changes only permitted repository paths. The evaluator resets a clean base checkout for each attempt, applies the patch once, and performs no manual repair. Build and test output are not returned to the model in the reported one-attempt setting. Any iterative repair experiment would be a separate condition with a separately declared feedback budget and must not be mixed with the results reported here.

\begin{figure*}[t]
\centering
\includegraphics[width=\textwidth]{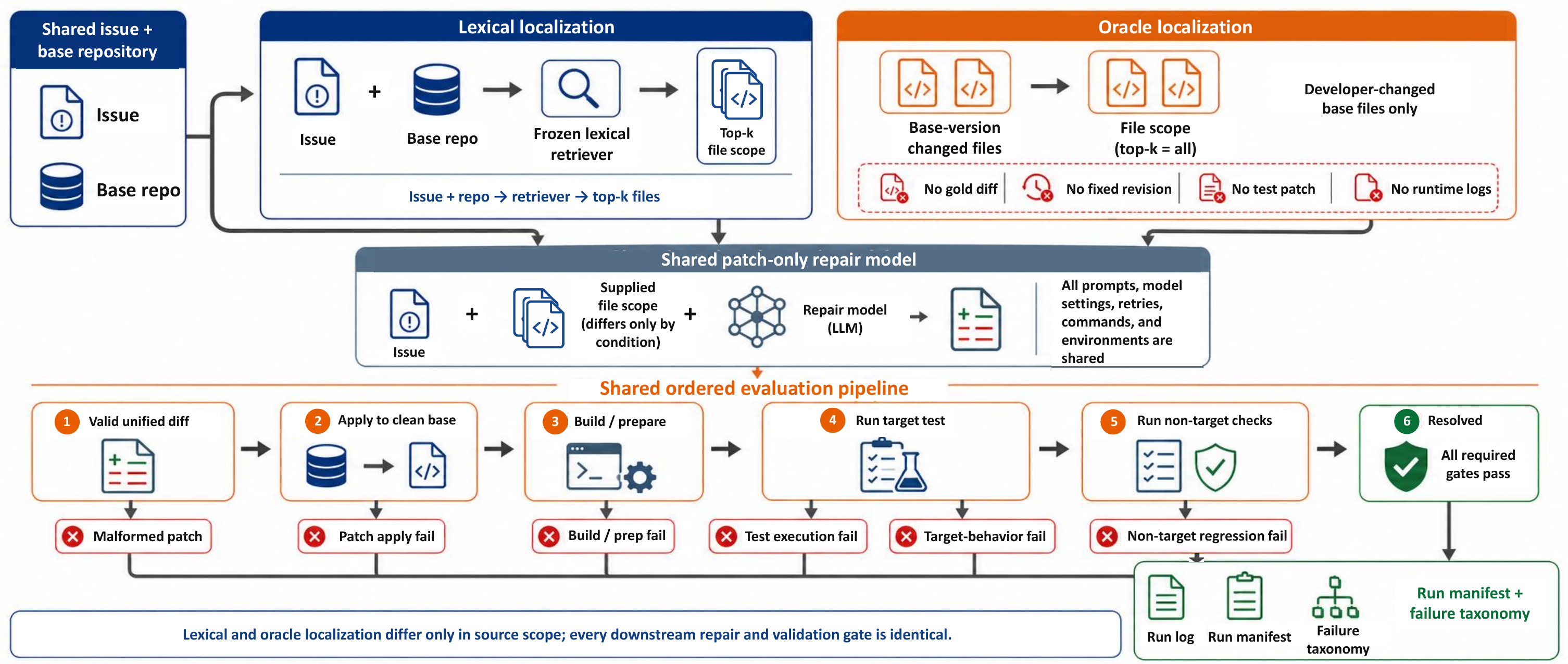}
\caption{Controlled repair evaluation.}
\label{fig:repair-evaluation-protocol}
\end{figure*}

Figure~\ref{fig:repair-evaluation-protocol} summarizes the source-scope conditions and ordered evaluator. Lexical and oracle localization change only the supplied files; every downstream generation and validation step remains fixed.

\subsection{Localization Conditions}

\textbf{Lexical localization} supplies the repair model with the top-$k$ files returned by a deterministic lexical retriever over the base revision using the issue text. The retriever indexes only files permitted by the instance policy. Before the full experiment, the release will freeze the tokenization, corpus filters, ranking function, $k$, tie-breaking rule, path normalization, and index checksum. The retriever must retain a per-instance ranked list so that Recall@$k$ can be recomputed without rerunning a mutable index.

Lexical localization is intentionally simple. It establishes a transparent end-to-end reference point and avoids using semantic embeddings, external documentation, or model-generated file selection as unreported sources of context. Issue-report-driven file ranking is a well-established bug-localization formulation \citep{zhou2012buglocalization,ye2014rankfiles}; our condition is deliberately narrower, ranking repository files from the issue text alone. Its purpose is not to claim that lexical retrieval is state of the art, but to quantify how much of the final repair gap is explained by source-file localization.

\textbf{Oracle localization} supplies the model with the base-version paths corresponding to files changed by the developer fix. It is an upper-bound condition for localization, not a realistic deployment setting. The model receives source contents from those paths but never receives the gold diff, fix commit, test patch, or reviewer evidence. The oracle path list is computed once from the frozen gold patch, normalized against the base checkout, and stored in the instance manifest. Comparing these two conditions separates the cost of localization from the model's ability to synthesize a repair once a relevant scope is available.

The oracle condition can still be difficult: a changed file may contain unrelated edits, a repair may require code outside the changed-file set, and the model receives no indication of the exact edited lines. Accordingly, oracle success is an upper bound for this particular file-scope protocol, not an estimate of an ideal repair system.

\subsection{Prompt Contract and Evaluator Isolation}

The shared repair prompt requests a patch only. It identifies the task, specifies the allowed source scope, requires a unified diff, and prohibits modifications to tests, environment recipes, or generated artifacts unless the benchmark task explicitly permits them. The six-model comparison uses the same prompt template and frozen source bundle for every model. The prompt and every included file are stored in the run manifest, so a later analysis can verify the model's actual information boundary.

Generation and evaluation are isolated processes. The generation service writes a candidate response and cannot execute repository commands. The evaluator consumes the normalized diff, performs patch application, invokes the containerized validator, and emits a structured outcome. Keeping the processes separate prevents hidden tool feedback from turning a nominal single-attempt repair into an interactive agent run. The same validators can then score gold patches, generated patches, and future baselines.

\subsection{Models and Controls}

We evaluate GPT-5.6-sol, GPT-5.6-terra-max, GLM-5.2, MiniMax-M3, Qwen3.8-Max, and Kimi-K3. Every model uses the same 168 instances, issue representation, repository snapshot, prompt template, source scope, patch format, retry policy, timeout, and validator. The final artifact will record model identifiers, API date, decoding parameters, context budget, concurrency, retry behavior, and per-instance cost and latency. Any model-specific protocol difference will be reported as a deviation.

The reported design uses one scored patch for each model-instance pair, producing a paired $6 \times 168$ matrix. Retries caused by an API transport failure are retained as infrastructure metadata and may repeat only the identical request before any model content is received. Retries after a model response, malformed patch, or evaluator outcome are forbidden in the one-attempt experiment.

\begin{table}[t]
\centering
\caption{Controls frozen before full repair evaluation.}
\label{tab:controlled-setup}
\begin{tabular}{p{0.37\linewidth}p{0.51\linewidth}}
\toprule
Control & Frozen record \\
\midrule
Model identity & Provider model string, API date, and request endpoint class. \\
Decoding & Temperature, top-p, maximum tokens, seed support, and stop rules. \\
Context & Prompt-template hash, source paths, byte or token budget, and truncation policy. \\
Execution & Container digest, validator revision, timeout, host architecture, and concurrency. \\
Attempt policy & One scored response; transport-only retry rule and complete retry log. \\
Analysis & Dataset manifest hash, retrieval-index hash, failure-classifier revision, and analysis-script commit. \\
\bottomrule
\end{tabular}
\end{table}

\subsection{Metrics and Failure Taxonomy}

The main comparison reports Apply, Repro, and Resolved on the 168 filtered instances. Apply measures whether the generated patch can be applied to the base revision. Repro records the issue-specific reproduction outcome, and Resolved records complete repair under the evaluator. Table~\ref{tab:main-results} reports only these three supplied metrics.

For the frozen eligible set $E$ and a metric indicator $s_i$, the reported rate is $\sum_{i \in E}s_i/|E|$. The same 168 instances form $E$ for every model and every reported metric. The set is fixed after the pre-run eligibility filter, and no model-specific post-run exclusion is allowed.

The evaluator retains detailed logs for later diagnosis, but unreported fields are not added to the main comparison.

Each unsuccessful attempt receives the first failed gate shown in Figure~\ref{fig:repair-evaluation-protocol} as its primary label, while raw logs retain secondary diagnoses. Preexisting infrastructure failures are excluded only by the frozen pre-run eligibility rule and remain in the release manifest.

The six-model comparison in Table~\ref{tab:main-results} fixes the supplied source scope and uses the same downstream generation, validation, and failure-classification steps for every model.

\section{Experimental Results}

\begin{table}[t]
\centering
\caption{Repair results on 168 filtered instances.}
\label{tab:main-results}
\small
\setlength{\tabcolsep}{3pt}
\begin{tabular}{@{}lrrr@{}}
\toprule
Model & Apply & Repro & Resolved \\
\midrule
GPT-5.6-sol & 100.0\% & 90.5\% & 50.6\% \\
GPT-5.6-terra-max & 99.4\% & 90.5\% & 48.8\% \\
GLM-5.2 & 100.0\% & 94.6\% & 57.7\% \\
MiniMax-M3 & 93.5\% & 88.7\% & 45.8\% \\
Qwen3.8-Max & 100.0\% & \textbf{96.4\%} & 48.8\% \\
Kimi-K3 & 100.0\% & 93.5\% & \textbf{66.7\%} \\
\bottomrule
\end{tabular}
\end{table}

\subsection{Overall Model Comparison}

Kimi-K3 obtains the highest Resolved score at 66.7\%, followed by GLM-5.2 at 57.7\% and GPT-5.6-sol at 50.6\%. GPT-5.6-terra-max and Qwen3.8-Max both reach 48.8\%, while MiniMax-M3 reaches 45.8\%. The ordering differs for Repro: Qwen3.8-Max leads at 96.4\%, followed by GLM-5.2 at 94.6\%. Complete resolution therefore changes the ranking obtained from Repro alone.

Kimi-K3 exceeds GLM-5.2 in Resolved by 8.9 percentage points. Qwen3.8-Max records the highest Repro score, yet its Resolved score is 17.9 points below Kimi-K3. MiniMax-M3 has the lowest Apply, Repro, and Resolved scores, while the other five models reach at least 99.4\% Apply.

\subsection{Metric-Level Analysis}

The three metrics in Table~\ref{tab:main-results} reveal different behavior. Apply spans 6.5 percentage points, Repro spans 7.7, and Resolved spans 20.8. Apply is nearly saturated for five models, but Resolved ranges from 45.8\% to 66.7\%. The run manifest retains the logs needed to diagnose the differences.

\section{Discussion}

OdinEval asks a narrow question: can a model restore one documented behavior with a patch that can be evaluated in the original environment? Kimi-K3 has the strongest Resolved result at 66.7\%, while Qwen3.8-Max leads Repro at 96.4\%. Their different rankings show that Repro is not the same as satisfying the full repair contract.

Metric choice therefore changes the apparent winner. Reporting only Apply would make four models indistinguishable, and reporting only Repro would favor Qwen3.8-Max. Resolved instead favors Kimi-K3 by a clear margin. OdinEval reports all three supplied metrics so that patch application, issue reproduction, and complete repair are not collapsed into one claim.

Generated tests cover issues without runnable developer tests, but same-model review is only a consistency check. Runtime validation remains mandatory, and one passing test does not prove full correctness. Frozen snapshots permit later comparisons if their information boundaries and budgets are disclosed.

The release also distinguishes a repeated evaluation from a new benchmark version. A new repair model can reuse the frozen issues, environments, tests, and validator. A localization study may change the supplied file scope while keeping all later gates fixed. Replacing a test, compiler, command, or acceptance rule changes the task itself and must produce a new version. This boundary prevents a reported gain from silently combining a better repair method with an easier oracle or environment.

\section{Threats to Validity}

\textbf{Construct validity.} Fail-to-pass tests may overfit symptoms or assumptions from the gold fix. Black-box review and two-state evidence reduce but do not remove this risk, so results separate developer, adapted, and generated tests. The reported comparison fixes one source scope and does not measure sensitivity to alternative localization methods.

\textbf{Internal validity.} Historical toolchains, validator changes, and model-service updates can alter outcomes. Manifests bind each result to its environment, validator, model, request, commands, and responses.

\textbf{External and conclusion validity.} The 168 filtered public issues may not represent all Odin development, and six models do not establish a universal ranking. We therefore bound the claims to the disclosed corpus, fixed denominator, and frozen evaluation protocol.

\section{Related Work}

Program repair has moved from search and symbolic reasoning \citep{le2012genprog,mechtaev2016angelix} to learned edits \citep{tufano2019empirical,lutellier2020coconut,jiang2021cure,zhu2021syntax} and LLM-based repair \citep{fan2023automated,jin2023inferfix,xia2023keep,yin2024thinkrepair}. OdinEval evaluates these approaches; it does not introduce another repair algorithm.

Repair benchmarks range from curated defects to issue-linked repositories \citep{just2014defects4j,widyasari2020bugsinpy,jimenez2024swebench}; QuixBugs pairs Java and Python defects \citep{lin2017quixbugs}. Unlike interactive systems such as SWE-agent \citep{yang2024sweagent}, OdinEval uses a patch-only protocol that keeps localization, generation, and validation separate.

SolEval and ArkRepoBench study repository-level generation and completion in Solidity and ArkTS \citep{peng2025soleval,wang2026arkrepobench}; HapRepair and ArkEval target OpenHarmony and ArkTS repair \citep{lin2025haprepair,xie2026arkeval}. OdinEval adds historical Odin toolchains, issue-specific oracles, and auditable fail-to-pass evidence. The tests support a repair claim but do not prove semantic correctness \citep{qi2015plausibility,barr2015oracle}.

\section{Conclusion}

OdinEval turns issue-linked Odin repairs into auditable benchmark instances. On 168 tasks, Kimi-K3 leads Resolved at 66.7\%, while Qwen3.8-Max leads Repro at 96.4\%. The ranking shift shows that Repro alone overstates complete repair. Frozen evidence makes this distinction reproducible.

{\small\bibliography{references}}
\end{document}